# Title: Revisiting time reversal and holography with spacetime transformations.

**Authors:**

Vincent Bacot[1], Matthieu Labousse[1,+], Antonin Eddi[2], Mathias Fink[1*], Emmanuel Fort[1*]

**Addresses:**

[1]Institut Langevin, ESPCI, CNRS, PSL Research University, 1 rue Jussieu, 75005, Paris, France.

[2]Laboratoire de Physique et Mécanique des Milieux Hétérogènes, ESPCI, CNRS, PSL Research University, 10 rue Vauquelin, 75005, Paris, France.

[+] Current address: Laboratoire Matériaux et Phénomènes Quantiques, Université Paris Diderot, Sorbonne Paris Cité, 10 rue A. Domon et L. Duquet, 75013 Paris, France.

[*] These authors contributed equally to this work.

**Abstract:**

Wave control is usually performed by spatially engineering the properties of a medium. Because time and space play similar roles in wave propagation, manipulating time boundaries provides a complementary approach. Here, we experimentally demonstrate the relevance of this concept by introducing instantaneous time mirrors. We show with water waves that a sudden change of the effective gravity generates time-reversed waves that refocus at the source. We generalize this concept for all kinds of waves introducing a universal framework which explains the effect of any time disruption on wave propagation. We show that sudden changes of the medium properties generate instant wave sources that emerge instantaneously from the entire space at the time disruption. The time-reversed waves originate from these "Cauchy sources" which are the counterpart of Huygens virtual sources on a time boundary. It allows us to revisit the holographic method and introduce a new approach for wave control.

Holographic methods are based on the time-reversal invariance of wave equations. They rely on the fact that any wave field can be completely determined within a volume by knowing the field (and its normal derivative) on any enclosing surface[1,2]. Hence, information reaching the 2D surface is sufficient to recover all information inside the whole volume. Based on these properties, Denis Gabor introduced the Holographic method, which provides an elegant way to back-propagate a monochromatic wave field and obtain 3D images. More recently, time-reversal mirrors exploited the same principles extended to a broadband spectrum to create time-reversed waves. This latter approach has been implemented with acoustic[3], elastic[4], electromagnetic[5] and water waves[6,7]. It requires the use of emitter-receptor antennas positioned on an arbitrary enclosing surface. The wave is recorded, digitized, stored, time-reversed and rebroadcasted by the antenna array. If the array intercepts the entire forward wave with a good spatial sampling, it generates a perfect backward-propagating copy. Note that this process is difficult to implement in optics[8,9], and the standard solution is to work with monochromatic light and use nonlinear regimes such as three-wave or four-wave mixing[10,11].

Here, within the general concept of spacetime transformations[12-16], we completely revisit the holographic method and introduce a new way to create wideband time-reversed wave fields in 2D or 3D by manipulating time boundaries. Time boundaries have recently received much attention because they have been shown to play a major role in several phenomena such as time refraction, dynamic Casimir effect, Hawking radiation, photon acceleration and self-phase modulation[17–25]. In addition, different suggestions to process wideband time-reversal have been proposed in optics to associate both time and spatial modulation of the medium refractive index. These suggestions, mainly for 1D propagation, rely on a dynamic tuning of photonic crystals[26–28].

Our approach is related to the Cauchy theorem, which states that the wave field evolution can be deduced from the knowledge of this wave field (and its time derivative) at one single time (the so-called initial conditions)[29]. It is the dual time equivalent of standard time reversal based on spatial boundaries. We use a sudden modification of the wave propagation properties of the medium to create a time-reversed wave. This time disruption realizes an instantaneous time mirror (ITM) in the entire space without the use of any antenna or memory. The information stored in the whole medium at one instant plays the role of a bank of memories.

We will subsequently introduce the concept of ITM and show its first experimental demonstration. The experiment is spectacular because it is conducted with water waves and can therefore be observed with the naked eye. We first interpret the backward wave propagation in ITM as an emission by isotropic sources created during the time disruption. These "Cauchy sources" define a new set of initial conditions for the wave field propagation after the ITM, allowing us to revisit the Huygens-Fresnel principle. We then discuss this experiment in terms of time discontinuities and conservation laws. Finally, we analyze the space-time symmetries of ITM compared to standard mirrors.

In the 19$^{th}$ century, Loschmidt challenged Boltzmann's attempt to describe irreversible macroscopic processes with reversible microscopic equations[30,31]. He imagined a daemon capable of instantaneously reversing all velocities of all particles in a gas. Such an operation can be ascribed to a change in initial conditions resulting in a time-reversed motion of all particles that would return to their initial positions. The extreme sensitivity to initial conditions that lies at the heart of chaotic phenomena in nonlinear dynamics renders any such particulate scheme impossible. Waves are more amenable because they can be described in many situations by a linear operator, and any error in initial conditions will not suffer from chaotic behavior. The wave analog of this

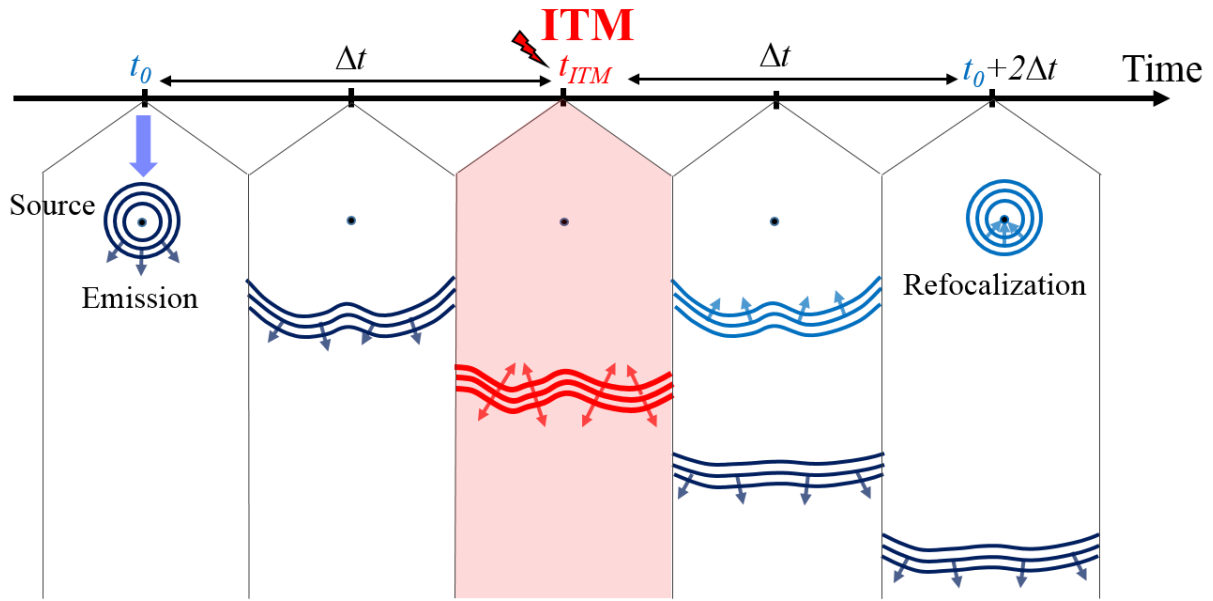

**Figure 1 | Schematic of the Instantaneous Time Mirror.** A wave source emits at time $t_0$ a wave packet which propagates in a given medium. A sudden spatially homogeneous disruption of the wave propagation properties occurs in the entire medium at time $t_{ITM} = t_0 + \Delta t$. It results in the production of a counter propagating time-reversed wave in addition to the initial forward propagating wave. The counter propagating wave refocuses at the source position at time $t_0 + 2\Delta t$.

Loschmidt daemon is related to the Cauchy theorem. The latter states that the future evolution of any wave field $\phi(\mathbf{r}, t)$ at position $\mathbf{r}$ and time $t$ can be inferred from the knowledge of the set of initial conditions $\left(\phi, \frac{\partial \phi}{\partial t}\right)_{t_m}$, with the field amplitude $\phi(\mathbf{r}, t_m)$ and time derivative $\frac{\partial \phi}{\partial t}(\mathbf{r}, t_m)$ at a given time $t_m$, in the whole space. The analog of the particle velocity reversal is to take new set of initial conditions $\left(\phi, -\frac{\partial \phi}{\partial t}\right)_{t_m}$ that cause a time-reversed wave whose time dependence is inverted. However, because of the wave superposition principle, the emergence of this time-reversed wave is not limited to this choice of initial conditions. For instance, the new initial condition $(\phi, 0)_{t_m}$ can be split into $\frac{1}{2}\left(\phi, \frac{\partial \phi}{\partial t}\right)_{t_m}$ associated with a forward wave and $\frac{1}{2}\left(\phi, -\frac{\partial \phi}{\partial t}\right)_{t_m}$ associated with a backward time-reversed wave. This particular choice erases the arrow of time by starting from a "frozen" picture of the wave field at time $t_m$ with no favored direction of propagation. Similarly, a new set of initial conditions $\left(0, \frac{\partial \phi}{\partial t}\right)_{t_m}$ in which the wave field is null would also comprise a backward-propagating wave with a negative sign. More generally, the superposition of backward- and forward-propagating waves results from the decoupling of the wave field from its time derivative (see Fig. 1 and the Supplementary Methods). Because both are bound together by the wave celerity, its disruption can lead to such decoupling. This offers a straightforward way to experimentally implement an instantaneous time reversal mirror.

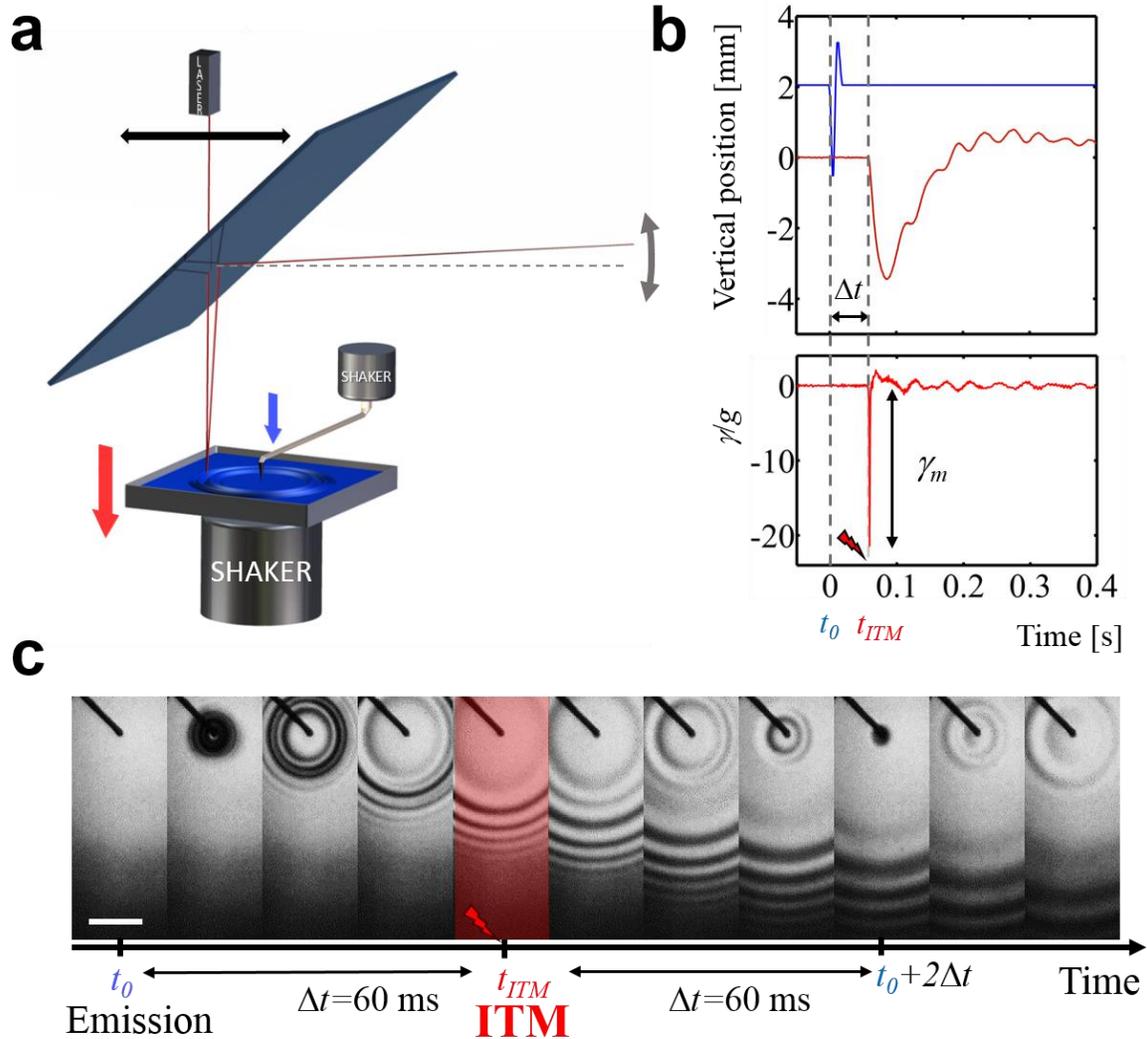

**Figure 2 | ITM experimental implementation. a**, Experimental setup. A bath of water is placed on a shaker in order to apply a vertical jolt. Another shaker is used to hit the water surface with a tip to generate surface waves. The deflection of a laser beam is used to measure the local surface slope. The laser is placed on a computer controlled translation device to scan the surface. **b**, Typical time variations of the vertical position of the emitter (a tip) in blue and of the bath together in red with bath acceleration $\gamma$ in an ITM experiment. $\gamma_m$ is the maximum downwards acceleration. $\Delta t$ is the time delay between the wave emission and the jolt. **c**, Image sequence of an ITM experiment (top view) with a point source showing the divergent wave and the time reversed wave which diverges again after focusing back at the source position. $\gamma_m = -21\ g$ and $\Delta t = 60$ ms. The scale bar is 1 cm (see Supplementary Video 1).

In this study, we use gravity-capillary waves to implement the concept of ITM. Because the surface wave celerity depends on the effective gravity, the disruption of the celerity is achieved by applying a vertical jolt to the whole liquid bath. Figure 2a shows the experimental setup. A bath of water is placed on a shaker to control its vertical motion. A plastic tip fixed on another shaker is used to hit the liquid surface and generate a point source of waves at time $t_0 = 0$. Figure 2b shows a typical time sequence of the vertical tip and bath motions used to generate the surface waves and implement the ITM. An image sequence of the wave propagation on the bath taken from above is shown in Fig. 2c. A circular wave packet centered on the impact point is emitted as the tip hits the surface. The average wave propagation velocity is on the order of magnitude of 10 cms$^{-1}$. After time $t_{ITM} = 60$ ms, a vertical downward jolt is applied to the bath. The bath acceleration reaches $\gamma_m = -21\,g$ in approximately 2 ms. The propagation of the initial outward-propagating wave is not significantly affected by this disruption. However, at the time of the disruption, we observe the apparition of a backward-converging circular wave packet that diverges again upon trespassing the original impact point source.

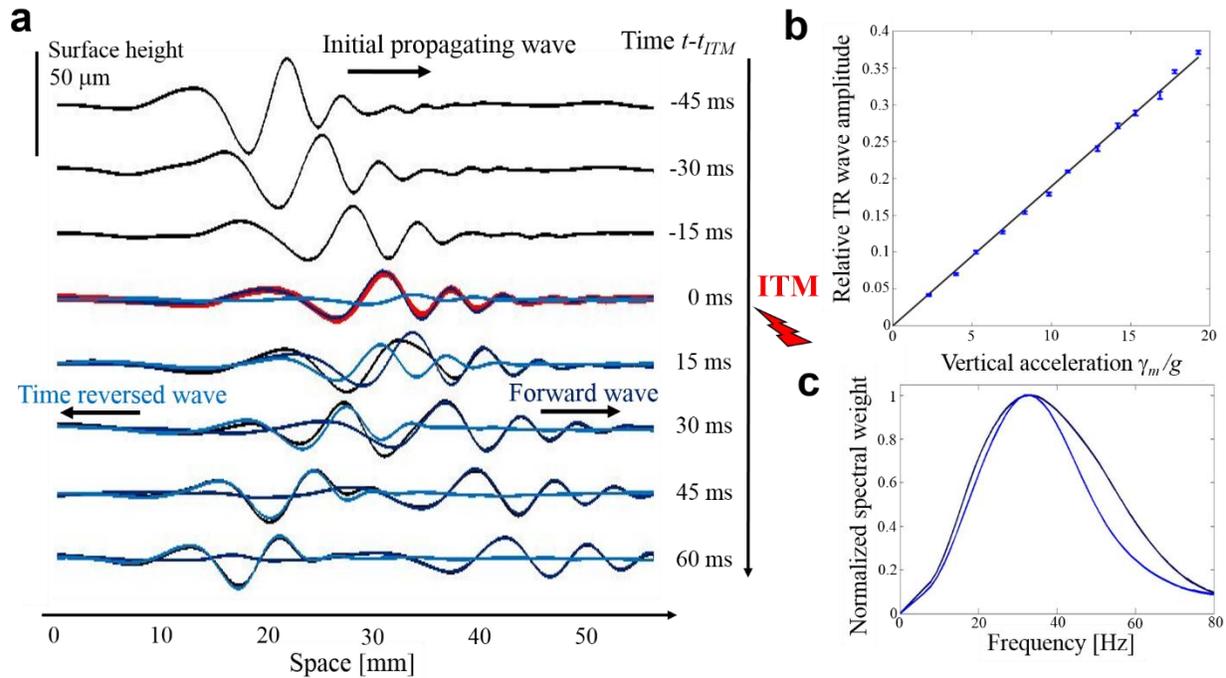

**Figure 3 | ITM on a wave packet. a**, Evolution of the profile of a wave packet produced by a point source and later subjected to an ITM (see Supplementary Video 2). The surface height (black solid line) is obtained by integrating slope measurements carried out on a line going through the emission point (see Fig. 1 and Supplementary Video 1). To the original wave packet propagating from left to right, a time reversed one propagating from right to left is added as the ITM occurs. The two counter propagating components of the surface profile are separated using Fourier analysis: The dark blue line represents the ongoing forward wave while the light blue line represents the time reversed wave. **b**, Relative Amplitude of the time reversed wave normalized by the forward wave amplitude as a function of the jolt amplitude. The measurement is performed in water at 1.6 cm from the point source. The ITM is applied with a time delay of $\Delta t = 170$ ms. The solid line is a linear fit which is coherent with the theory (see Supplementary Methods). **c**, Normalized spectra of the time reversed wave packet (light blue) and of the initial forward wave packet (dark blue). Both are similar with respective maximum frequency $\omega_{max} \approx 35$ Hz and full width at half maximum $\Delta\omega \approx 35$ Hz.

Figure 3a is a time sequence of the profile of a wave packet propagating originally from left to right. The wavelength spreading induced by dispersion is clearly visible. The ITM generates a time-reversed wave packet propagating in the opposite direction. The resulting surface profile can be decomposed into the two counter-propagating wave packets using Fourier analysis. We observe that the shape of the backward wave packet is very similar to that of the initial wave packet. Both profiles almost superimpose in shape and position when measured at identical times $\Delta t$ from the ITM. A phase shift of approximately $\pi/2$ is observed between the forward and backward wave packets at the time of the ITM. In contrast with standard reflection, the backward wave packet is not spatially reversed. The time-reversed nature of the backward wave allows the wave packet to compensate for dispersion. The fast short wavelengths will catch up with the slow long wavelengths, thus refocusing the wave packet. Its amplitude linearly depends on the vertical acceleration of the bath (Fig. 3b). ITM is a broadband time reversal mirror. The time-reversed spectrum is independent of the jolt amplitude and is nearly identical to that of the initial wave (Fig. 3c).

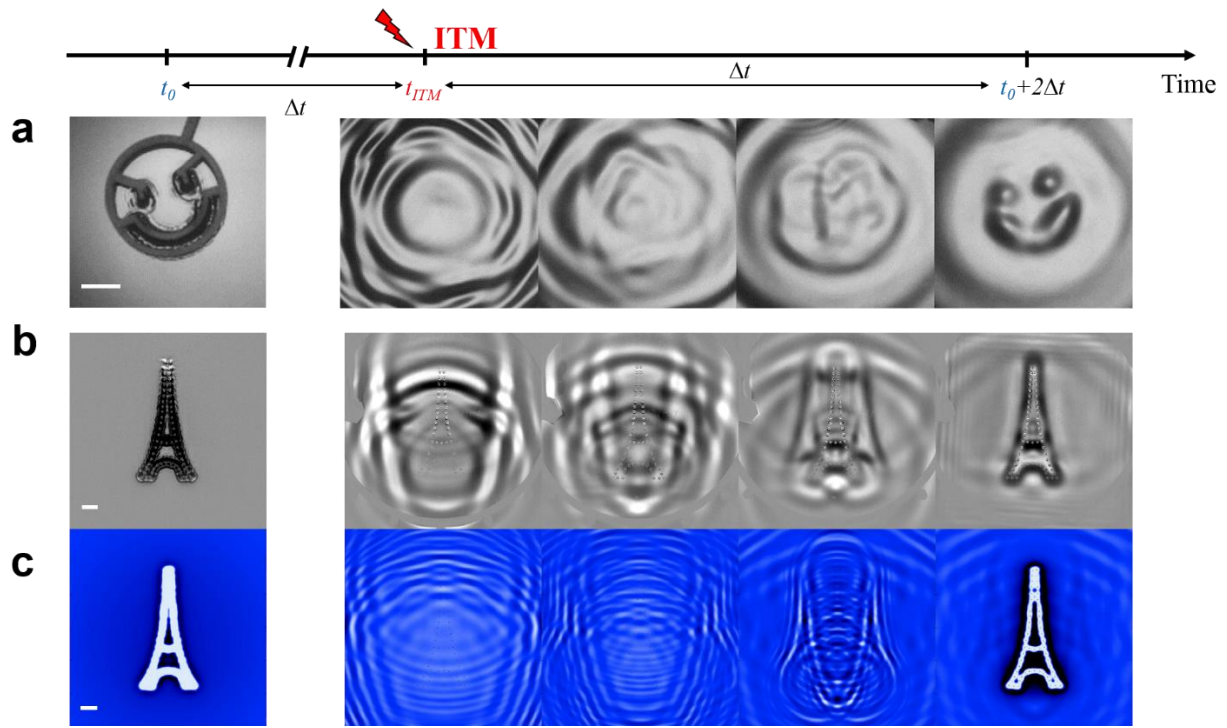

**Figure 4 | Image sequence of the instantaneous time reversal of a complex wave field.** The source is composed of (**a**) tips that hit the surface positioned in the shape of a Smiley (see Supplementary Video 3) or (**b**) air blowing between two sealed Plexiglas plates placed at 1 cm above the bath with holes positioned in the shape of an Eiffel tower (see Supplementary Video 4). In the sequence (**b**), the image without blowing has been subtracted as a reference. **c**, is the numerical simulation of (**b**) using the ITM model for water waves and the experimental jolt profile (see Supplementary Method and Video 5). The images on the left show this emission process. At the instant of ITM, the wave field features a complex interference pattern in which the original shape is not apparent anymore. As the time reversed wave refocuses, the shape of the source becomes visible again. Time interval between two images is 26 ms for (**a**) and 66 ms for (**b**) and (**c**). The scale bar is 1 cm.

Figure 4 shows two examples of ITM performed on sources with complex source shapes. In both cases, the ITM disruption occurs long after the wave field has lost any resemblance to its initial shape at the time of emission. The refocus back to its initial shape indicates the time reversal nature of the process.

We now focus on the underlying principles of ITM. ITM is implemented through a wave celerity disruption induced by the gravity jolt. For the sake of generality, let us consider waves governed by d'Alembert's wave equation. We introduce a time-dependent phase velocity $c(t) = c_0/n(t)$, where $n(t)$ is a time-dependent index and $c_0$ is the phase velocity in the absence of ITM. The disruption undergone by the medium in an ITM can be modelled by a $\delta$–Dirac function such that $c(t)^2 = c_0^2(1 + \alpha\delta(t - t_{ITM}))$. The wave equation can be written as a nonhomogeneous equation in which the equivalent source term $s(\boldsymbol{r},t)$ is induced by the velocity disruption (see Supplementary Methods):

$$\Delta\phi(\boldsymbol{r},t) - \frac{1}{c_0^2}\frac{\partial^2\phi}{\partial t^2}(\boldsymbol{r},t) = s(\boldsymbol{r},t), \qquad (1)$$

with $s(\boldsymbol{r},t) = -\frac{\alpha}{c_0^2}\delta(t - t_{ITM})\frac{\partial^2\phi}{\partial t^2}(\boldsymbol{r},t)$.

The source term is localized in time but delocalized in space. It corresponds to an instantaneous source that is proportional to the second time derivative of the wave field at the instant $t_{ITM}$ of the disruption. Equation (1) can also be applied in the Fourier domain to water waves to take care of dispersion. All the results subsequently presented for d'Alembert waves can thus be similarly recovered for water waves (see Supplementary Methods). Considering both the specific dispersion relation of these waves and the experimental profile of the jolt, we used Equation (1) to simulate ITM action in our experiments (see Supplementary Methods).

This description with a source term allows us to revisit the Huygens-Fresnel theory. To model the wave propagation, Huygens hypothesized that every point on a wavefront emits secondary spherical wavelets[28]. The wavefront at any later time $t + \Delta t$ conforms to the upper envelope of the wavelets emanating from every point on the wavefront at a prior instant $t$ (Fig. 5a). However, neglecting the backward-propagating envelope was arbitrary. Only later, Fresnel[30], followed by Kirchhoff[31], ensured that the wavelets interfere destructively in the backward direction and maintain the expected forward propagation by adding a dipolar component to the secondary sources.

In our experiment, the temporal disruption modifies the classical interplay between the dipolar and monopolar sources that causes a propagative wave. It suddenly creates *real monopolar sources* $s(\boldsymbol{r}, t)$ instantaneously in the whole space (see Eq. 1). These sources isotropically radiate, generating an additional wave field, both forward and backward (see Fig. 5a). Because they modify the initial conditions of the wave field on a time boundary, these sources can be termed *Cauchy sources*.

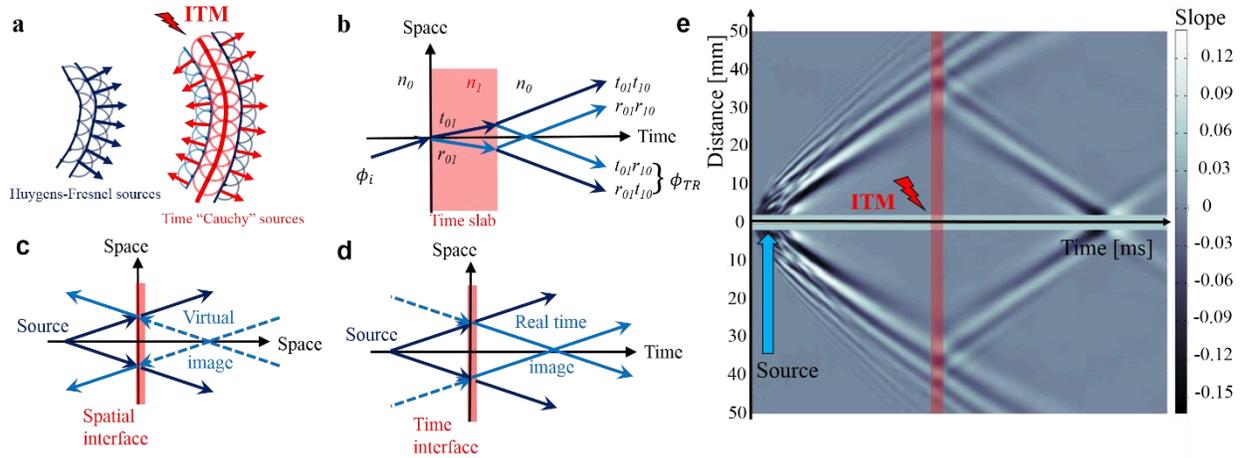

**Figure 5 | The time equivalent of a mirror. a**, Schematics comparing the standard wave propagation using secondary Huygens-Fresnel sources and the creation of real secondary "Cauchy" sources during ITM. **b**, Schematics of an incident wave field $\phi_i$ impinging on a time slab. $\phi_B$ is the backward propagating field. $t_{ij}$ and $r_{ij}$ are the transmission and reflection time Fresnel coefficients between medium $i$ and $j$ (see Supplementary Methods). Schematics of a standard spatial mirror (**c**) and an instantaneous time mirror (**d**). These mirrors are the limit case of an infinitely thin space slab and an infinitely thin time slab respectively (see Supplementary Methods). The directions of the arrows show the spatio-temporal directions of wave propagation. With a standard mirror, the reflected beams are equivalent to those emitted by a virtual source situated at the symmetric point on the other side of the mirror. For the ITM, the image is situated at the symmetric instant of that of emission with respect to the time interface. Because of causality, this image is real and corresponds to a focalization point. **e**, Spatiotemporal diagram of the slope measurements in an ITM experiment with a point source (see Fig. 2c). The diagram is symmetrized for clarity. The area in red highlights the ITM disruption occurring 80 ms with an approximate width of 8 ms. From this moment on, the time reversed wave is observed. As it converges back to the emission point, this packet narrows down; the effect of dispersion being compensated.

What is the relation between these Cauchy sources and the change of initial conditions induced by ITM? Just before the ITM at $t_{ITM}^-$, the wave field is associated with $\left(\phi, \frac{\partial \phi}{\partial t}\right)_{t_{ITM}^-}$. It is modified by the disruption into $\left(\phi, \frac{\partial \phi}{\partial t}\right)_{t_{ITM}^+}$ just after the ITM at $t_{ITM}^+$. The new initial state is given by (see Supplementary Methods):

$$\left(\phi, \frac{\partial \phi}{\partial t}\right)_{t_{ITM}^+} = \left(\phi(\boldsymbol{r}, t_{ITM}^-), \frac{\partial \phi}{\partial t}(\boldsymbol{r}, t_{ITM}^-) + \frac{\alpha}{c_0^2}\frac{\partial^2 \phi}{\partial t^2}(\boldsymbol{r}, t_{ITM}^-)\right) \quad (2)$$

This new initial state can be decomposed into the superposition of the original state of the unperturbed wave field $\left(\phi, \frac{\partial \phi}{\partial t}\right)_{t_{ITM}^-}$ plus an added state $\left(0, \frac{\alpha}{c_0^2}\frac{\partial^2 \phi}{\partial t^2}\right)_{t_{ITM}^-}$. This latter term can again be decomposed in two states as previously discussed by using the superposition principle: $\frac{\alpha}{2c_0^2}\left(\frac{\partial \phi}{\partial t}, \frac{\partial^2 \phi}{\partial t^2}\right)_{t_{ITM}^-}$ and $-\frac{\alpha}{2c_0^2}\left(\frac{\partial \phi}{\partial t}, -\frac{\partial^2 \phi}{\partial t^2}\right)_{t_{ITM}^-}$, which correspond to a forward-propagating wave field and a time-reversed backward-propagating wave field, respectively. Both wave fields are proportional to the time derivative of the original incident wave field. Provided that the bandwidth of the time-reversed wave is not too large compared to the central frequency, these wave fields are proportional to the original wave field itself as observed in the experiments. Note that the expected π/2 phase shift between the wave field and its derivative is the one observed in the experiment (see Fig. 3a). In practice, the time-reversed bandwidth is limited by that of the ITM disruption (see Fig. 3c), which should be non-adiabatic for wave propagation.

ITM can be analyzed in the framework of time refraction[16-19]. The instantaneous time disruption for the wave speed can be considered as the limiting case of a rectangular time profile with two discontinuities: at time $t_{ITM}^-$, the wave speed jumps from $c_0$ to $c_1 = c_0/n_1$ and then, at time $t_{ITM}^+$, changes back to its original value $c_0$. A temporal discontinuity in a homogeneous medium conserves the momentum but not the energy. In our experiment, this energy brought to the wave field is provided by the jolt. The time analog of the Fresnel formula can be obtained from conservation laws[18-19]. Hence, a monochromatic wave $e^{i(\mathbf{k}.\mathbf{r}-\omega_0 t)}$ of wave vector $\mathbf{k}$ and angular frequency $\omega_0$ is split at the time discontinuity in a "transmitted" wave $t_{01}e^{i(\mathbf{k}.\mathbf{r}-\omega_1 t)}$ and a 'reflected' wave $r_{01}e^{i(\mathbf{k}.\mathbf{r}+\omega_1 t)}$, where $\omega_1 = \omega_0/n_1$ is the angular frequency in medium 1, and $t_{01}$ and $r_{01}$ are temporal Fresnel coefficients for time refraction and reflection, respectively. Each wave emerging from the first temporal discontinuity will be split again into two waves at the second discontinuity (see Fig. 5b). This time slab is the time analogue of a Fabry-Pérot resonator. However, because of causality, multiple reflections are not permitted[17,18]. The time-reversed wave field is thus the result of interference between two backward waves with opposite signs (because $r_{01}t_{10} = -r_{10}t_{01}$). This explains why the resulting time-reversed field is not the perfect time reverse of the incident wave field $\phi$, but rather of its derivative $\partial\phi/\partial t$ in the limiting case of an instant disruption (see the Supplementary Methods).

We now focus on the spatio-temporal symmetries of ITMs using plane waves without loss of generality. A standard "spatial" mirror (see schematic Fig. 5c) changes the sign of the wave vector normal component to its surface $k_\perp$. It changes the wave characteristics as $(k_\perp, k_\parallel, \omega) \rightarrow (-k_\perp, k_\parallel, \omega)$ by performing a space reversal in the normal direction. The incident wave $\phi(x, ..., t)$ becomes the reflected wave $\phi(2x_m - x, ..., t)$ for a mirror positioned along the $x$ axis at $x_m$. For a point source, the reflected waves appear as though it is emitted from a virtual image located on the other side of the mirror. As previously mentioned, ITM symmetry is given by $(k_\perp, k_\parallel, \omega) \rightarrow (k_\perp, k_\parallel, -\omega)$. This corresponds to a time reversal: the incident wave $\phi(x, ..., t)$ becomes the time-reversed wave $\phi(x, ..., 2t_{ITM} - t)$ for an ITM at $t_{ITM}$. The direction of propagation for a plane wave is given by its phase $\mathbf{k}.\mathbf{r} - \omega t$, which depends on the relative signs of $\mathbf{k}$ components and $\omega$. Hence, in terms of symmetries, an ITM is equivalent to $(k_\perp, k_\parallel, \omega) \to (-k_\perp, -k_\parallel, \omega)$. Because all components of **k** are reversed, waves propagate backward. In the space-time representation of the ITM (see Fig. 5d), the waves refocus at their emitter positions but on the other time side of the mirror in the time domain. This can be observed directly in the experiment on the spatio-temporal graph of waves emitted from the point source undergoing an ITM (see Fig. 5e). Note that this transformation is also directly related to negative index materials. Time reversal and negative refraction have been shown to be intimately linked processes[26].

Manipulating the wave propagation from the time boundaries offers a new approach to control and manipulate wave propagation. Time disruptions perform instantaneous time mirrors simultaneously acting in the entire space at once and without the use of external emitters. This approach will be generalized to create a dynamic control of the spatio-temporal boundaries of the medium. Water waves present unique advantages for implementation and visualization. In this perspective, several possibilities of precise and rapid spatio-temporal wave control are offered, for instance, by using ultrasound or electrostatic forces on the liquid surface. In the future, we intend to use these concepts of spacetime transformation to perform water wave time cloaking and to revisit Faraday instability as a periodic time Bragg mirror. From this new perspective, we will experimentally address fundamental issues such as the dynamic Casimir effect.

**Acknowledgments:**

We are grateful to Yves Couder and Stéphane Perrard for fruitful and stimulating discussions. We thank Abdelhak Souilah and Xavier Benoit-Gonin for their help in building the experimental set-up. The authors acknowledge the support of the AXA research fund and LABEX WIFI (Laboratory of Excellence ANR-10-LABX-24) within the French Program "Investments for the Future" under reference ANR-10- IDEX-0001-02 PSL* .

**Author Contributions Statement**

All the authors discussed, interpreted the results and conceived the theoretical framework. M. F. and E. F. conceived the initial concept. V. B., A. E., M. F. and E. F. designed the experiment. V. B. and A. E. performed the experiments. M. L. extended the model to water waves and designed the simulations. V. B., M. F. and E. F. wrote the paper. All authors reviewed the manuscript. Correspondence and requests for materials should be addressed to mathias.fink@espci.fr and emmanuel.fort@espci.fr

**Competing financial interests**

Authors declare they have no competing financial interests